\documentclass{appolb}
\usepackage{epsfig}
\begin{document}
\title{Recent results from the Tevatron}
\author{C. Royon
\address{DAPNIA/Service de physique des particules, CEA/Saclay, 91191 
Gif-sur-Yvette cedex, France\\and Fermi National Accelerator Laboratory, 
Batavia, Illinois 60510, USA}}
\maketitle
\begin{abstract}

In these lectures, we describe some recent results from the D\O\ and CDF experiments at
the Tevatron.

\end{abstract}
\section{Introduction and description of D\O\ and CDF experiments}
In this article, we will describe some of the newest results obtained 
by the D\O\ and CDF experiments at the Tevatron in 2005. We will give in turn
some results about QCD, top, b physics, new phenomena and prospects for Higgs
boson searches.

The Tevatron is a $p \bar{p}$ collider located near Chicago with a
center-of-mass energy of 1.96 TeV, which is the highest energetic machine before
the start of the LHC. The expected sensitivity to physics beyond the standard
model is thus high. The two main experiments (D\O\ and CDF) are installed along
the ring and provide independent physics analyses to allow cross checks between
the results.

The accumulated luminosity \footnote{The luminosity is directly related to the
number of events which have been taken by the experiment, since $N = \sigma
\times {\cal L}$ where N, $\sigma$, and ${\cal L}$ are respectively the number
of events for a given process, the cross section for that process and the
luminosity.} by the D\O\ experiment is given in Fig. 1 until the time of the
Summer school. The expected luminosity before 2009 when the Tevatron will
probably be turned off is expected to be between 4 and 8 fb$^{-1}$. 
The luminosity accumulated by the CDF experiment is found to be similar and
slightly higher. The data taking efficiency, which gives the percentage of time
when D\O\ is able to take data, is noticeably well above 90\%.

A scheme of the D\O\ detector is given in Fig. 2. We will give the
description of the D\O\ detector starting from the center to the outside
\cite{nimd0}. The
most central part comprises a (forward and central) silicon and a fiber tracking
detector, which allows to measure precisely the location and momentum of charged
particles. The tracking detector is surrounded by a solenoid which delivers a
magnetic field of 2 T. The compensating, finely segmented, liquid argon 
and uranium calorimeter provides nearly a full solid angle coverage up to
a rapidity larger than 4. The muon detector is composed of the central muon
proportional drift tubes, scintillating detectors used in the trigger,
and mini-drift tubes in the forward region, allowing a muon detection up to a
rapidity of 2. A toroid magnet allows to reconstruct the muon momentum using
the muon system only, and a better resolution is obtained by combining this
information with the ones from the tracking detectors. The CDF
detector has similar performances and is composed of a central
tracking and silicon detector, a calorimeter made of lead sheets sandwiched with
scintillator for the electromagnetic part, and of iron plates and scintillator
for the hadronic part, and a muon detector. The leveir arm for the tracking
detector is larger than for D\O\ because of the space availability (we recall
that D\O\ did not have any central magnet in Run I). 

We will now describe the different physics results in turn.

\begin{figure}
\begin{center}
\epsfig{file=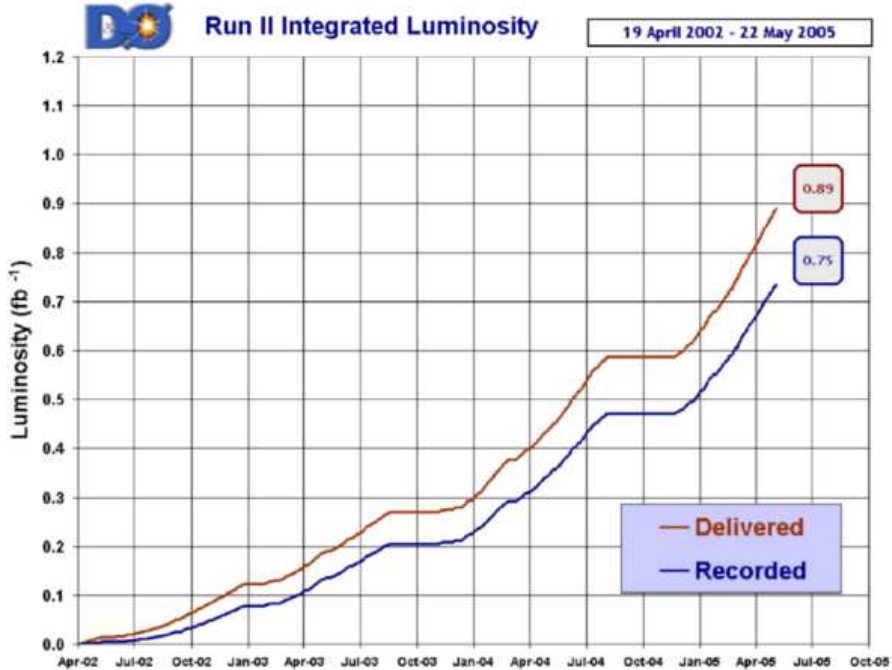,width=12cm}
\caption{Integrated luminosity accumulated by the D\O\ experiment.}
\end{center}
\label{fig1}
\end{figure}

\begin{figure}
\begin{center}
\epsfig{file=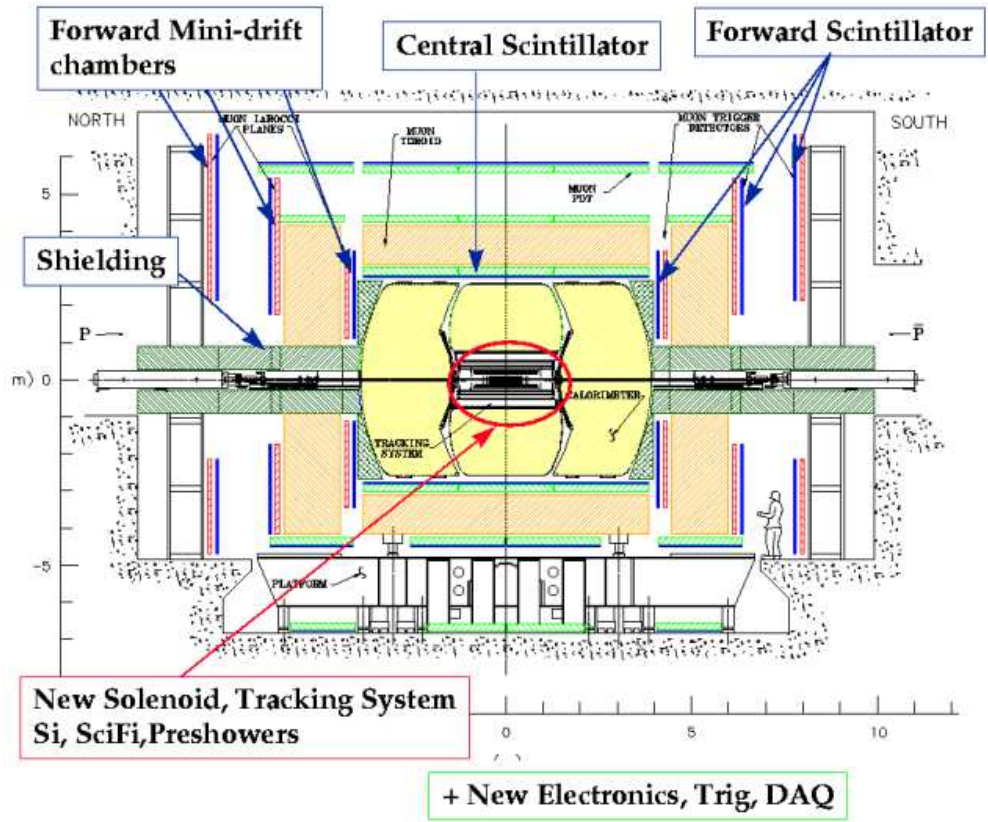,width=12cm, angle=270}
\caption{Scheme of the D\O\ detector.}
\end{center}
\label{fig2}
\end{figure}

\section{Results on QCD}

\subsection{Why measuring the QCD cross sections at the Tevatron?}
In this paragraph, we will discuss the CDF and D\O\ results on QCD. First, it is
useful to notice that these experiments lead to results
quite complementary to the ones from HERA, and
the previous fixed target experiments. As shown in Fig. 3, the
kinematical plane in ($x$, $Q^2$) ($x$ is the proton momentum fraction carried
by the interacting quark, and $Q^2$ is the squared energy transferred at the 
lepton vertex) reached at HERA extends noticeably the reach of the previous
fixed target experiment. The Tevatron experiments are also sensitive to higher
$Q^2$ and higher $x$ value. The constraint on the gluon density at high $x$ in
particular is coming mainly from the Tevatron and fixed targets experiments.
In that sense, the data taken at HERA and Tevatron are complementary to obtain
precisely the quark and gluon densities from Dokshitzer Gribov Lipatov Altarelli
Parisi (DGLAP) QCD fits \cite{dglap}. The $F_2$ structure function measurements
as well as the QCD fits are given in Ref. \cite{h1zeusf2}. The uncertainty on
the gluon distribution at high $x$ is large and reaches more than 50\% for $x$
larger than 0.5. 

\begin{figure}
\begin{center}
\epsfig{file=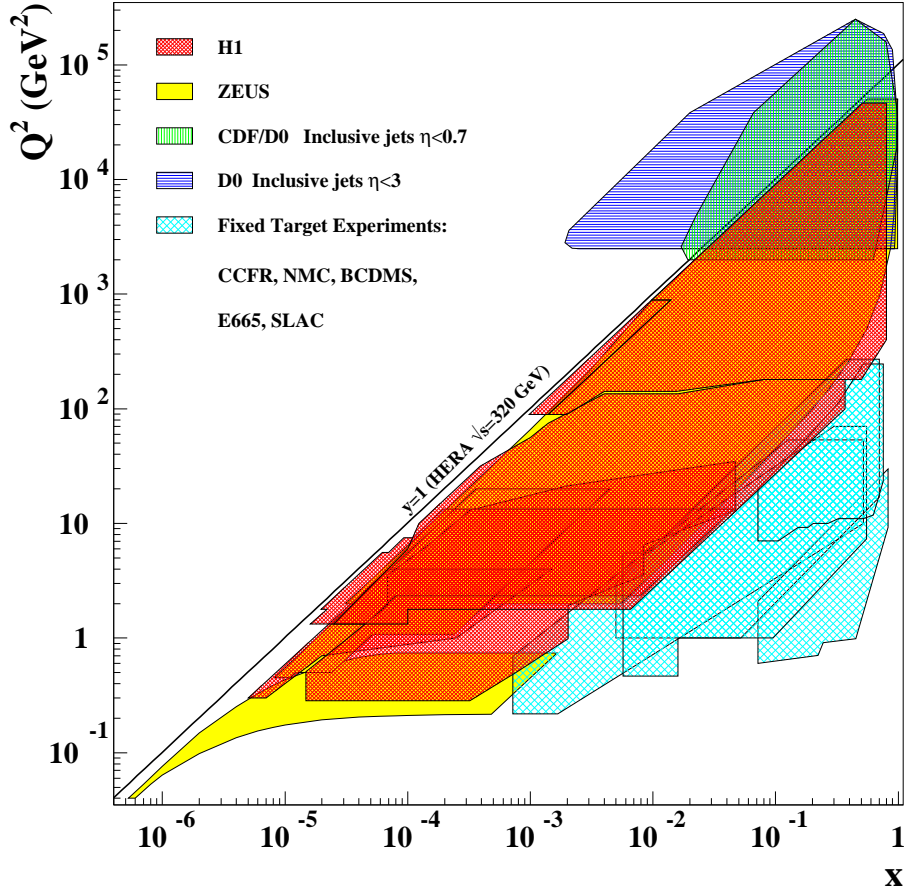,width=12cm}
\caption{Kinematic plane in ($x$, $Q^2$) reached by Tevatron, HERA and the fixed
target experiments.}
\end{center}
\label{fig3}
\end{figure}

\subsection{QCD inclusive jet cross section measurements}
The CDF and D\O\ experiments performed a preliminary measurement of the
inclusive jet cross sections as a function of their transverse momentum to probe
the high-$x$ gluon density. The preliminary measurement performed by the D\O\
collaboration with a luminosity of about 378 pb$^{-1}$ and two bins in rapidity
is given in Fig. 4. The measurement in the lowest bin in rapidity
($|y|<0.4$) has been multiplied by 10 to be able to distinguish between both
measurements. The data are compared with NLO calculations using the CTEQ6.1M
parametrisation \cite{cteq} and the NLOJET++ program \cite{nlojet}. There is a
good agreement between the measurement and the QCD calculation over 9 orders
of magnitude. The data over theory plot for the same data is given in Fig.
5. The data points are in black for both rapidity bins and the
systematic uncertainties are indicated by the yellow band. The systematics are
largely dominated by the uncertainty on jet energy scale. The jet energy scale
is determined using the $p_T$ balance in photon and jet events where the
electromagnetic energy scale is known using $Z$ decaying into $e^+e^-$, and the
photon and the jet are required to be back-to-back. The theory corresponds to
NLO QCD calculations using the CTEQ6.1M parametrisation. The CTEQ6.1 parton
distribution uncertainty (mainly due to the bad knowledge of the gluon density
at high $x$) is given by the red dashed line, and the difference with the
MRST2004 \cite{mrs} parametrisation by the blue dotted line. The
present uncertainties of the measurement do not allow a further constraint of
the parton distribution. A significant improvement of the jet energy scale
uncertainty is expected in the beginning of 2006 which will allow to constrain
the high-$x$ gluon density. Let us also note that a measurement at higher
rapidity is also another way to be sensitive to the high-$x$ gluon since pure
gluon-gluon and quark-gluon jets are more present at higher rapidity than
quark-quark processes. A preliminary measurement with a lower luminosity has
already performed at lower luminosity and is being redone \cite{d0jetold}.
The measurement of the dijet mass cross section has also been performed by the
D\O\ collaboration and will allow to put some new limits on compositeness in the
near future since this measurement is sensitive to possible quark or gluon
substructures \cite{d0jetold}. The CDF collaboration performed a similar 
measurement of the inclusive jet $p_T$ cross section using the $k_T$ algorithm
\cite{jetcrosscdf}.

\begin{figure}
\begin{center}
\epsfig{file=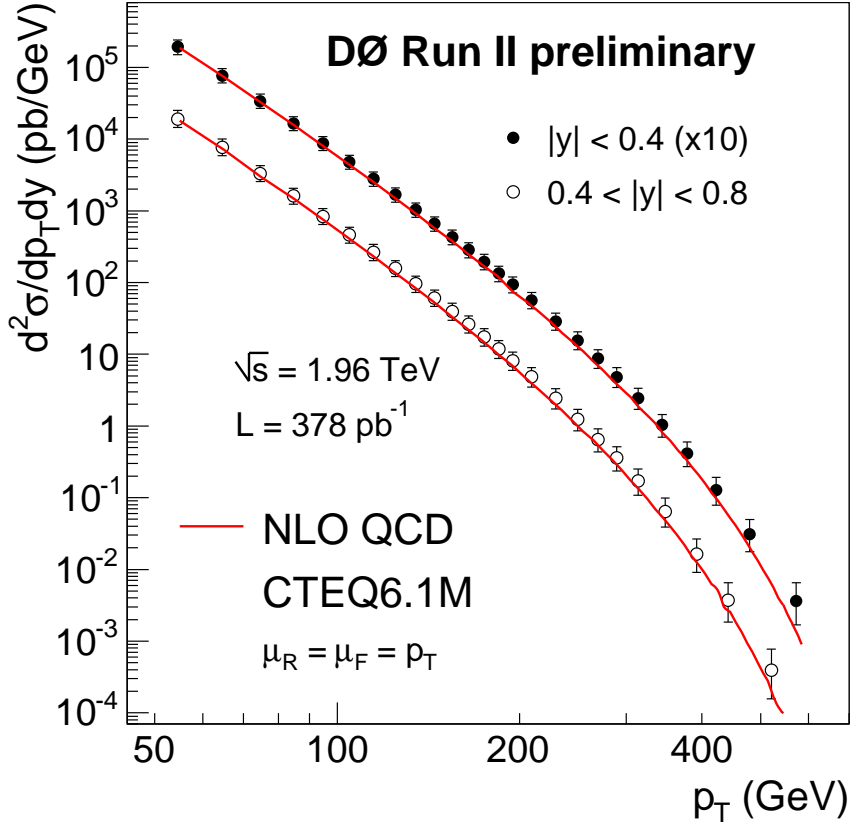,width=12cm}
\caption{Measurement of the inclusive jet cross section as a function of
their transverse momentum from the D\O\ collaboration for two bins in rapidity.}
\end{center}
\label{fig4}
\end{figure}

\begin{figure}
\begin{center}
\epsfig{file=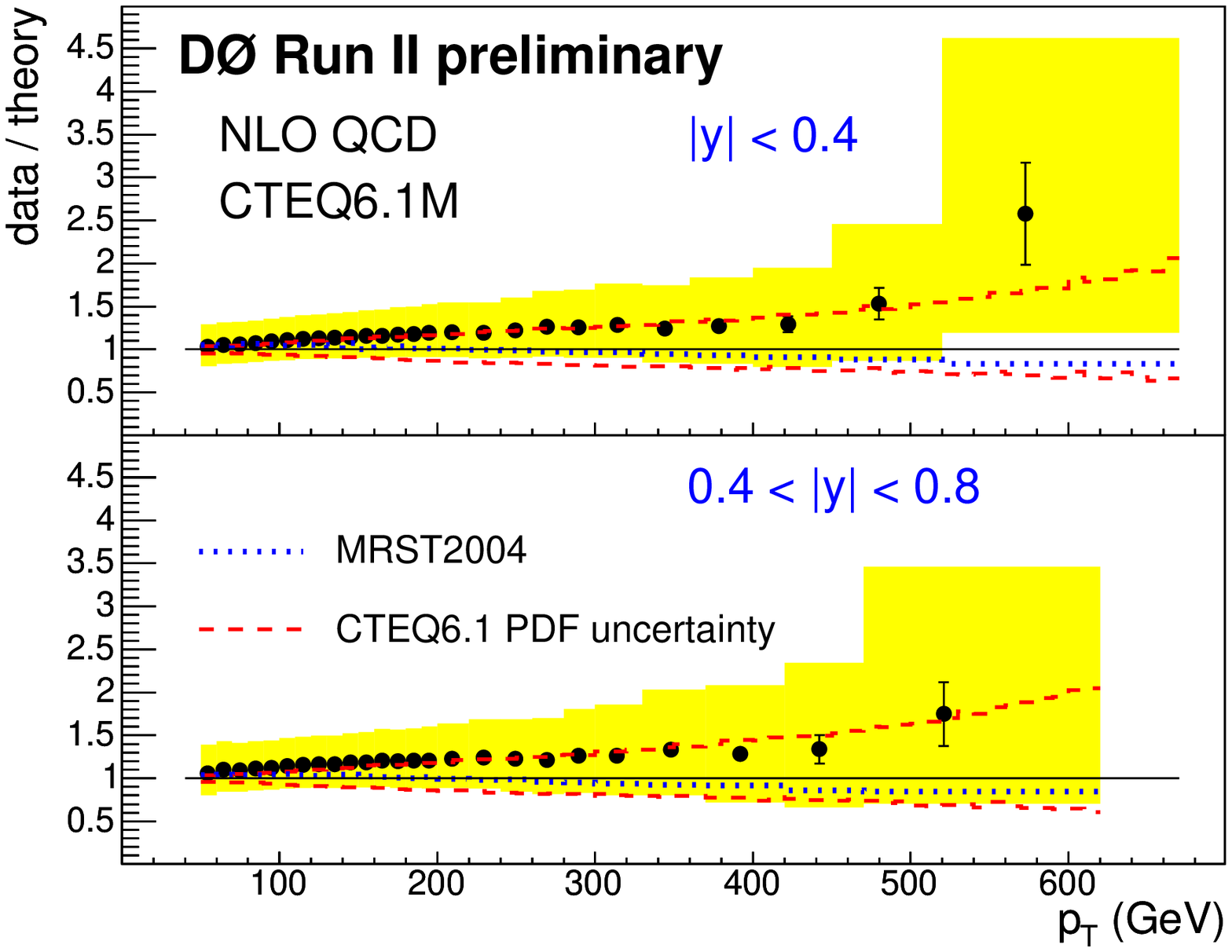,width=12cm}
\caption{Data over theory ratio for the inclusive jet cross section measurement
from the D\O\ experiment.}
\end{center}
\label{fig5}
\end{figure}

\subsection{Measurement of the difference in azimuthal angles between jets}
Another measurement which has been performed by the D\O\ collaboration is the 
measurement of the difference in azimuthal angle between the two leading jets in
QCD events \cite{d0deltaphi}. The azimuthal angle between the two leading jets
is expected to be close to $\pi$ for pure dijet events whereas the angle will be
less than $\pi$ in the case of multiple jet events. The angle measurement is
thus directly sensitive to higher order effects without measuring effectively
the jet structure of the event. Furthermore, this measurement does not suffer
too much from the jet energy uncertainty due to jet energy scale since it
depends on angles and not directly on energy. The measurement of
the relative differential cross section in azimuthal angle is shown in Fig.
6 in four different bins in jet transverse energy. The measurement is
compared to LO and NLO calculation in dashed and full lines respectively. We
notice a disagreement at low values of $\Delta \Phi$ with the LO calculation
since the number of multijet events is too small at LO. NLO calculation agrees
nicely with the data except at very large $\Delta \Phi$ close to $\pi$ where not
enough soft radiation is produced. We also show the sensitivity of this
measurement on Monte Carlo tuning in Fig. 7. The HERWIG \cite{herwig}  
Monte Carlo shows a good agreement with data, whereas the default PYTHIA
\cite{pythia} shows some discrepancy. Increasing initial state radiation in
PYTHIA (technically, PARP(67) was increased from 1. to 4.) solves the problem, and the sensitivity on this
parameter is shown in Fig. 7 by the blue band. It is quite important to
determine precisely the parton distributions in the proton and to tune the
existing Monte Carlo to be able to obtain precise predictions at the LHC, which
is fundamental to see some effects beyond the Standard Model, especially in the
jet channels. We can quote in particular the importance of understanding the jet
cross sections for $R$-parity violated SUSY or the search for higher dimensions.

\begin{figure}
\begin{center}
\epsfig{file=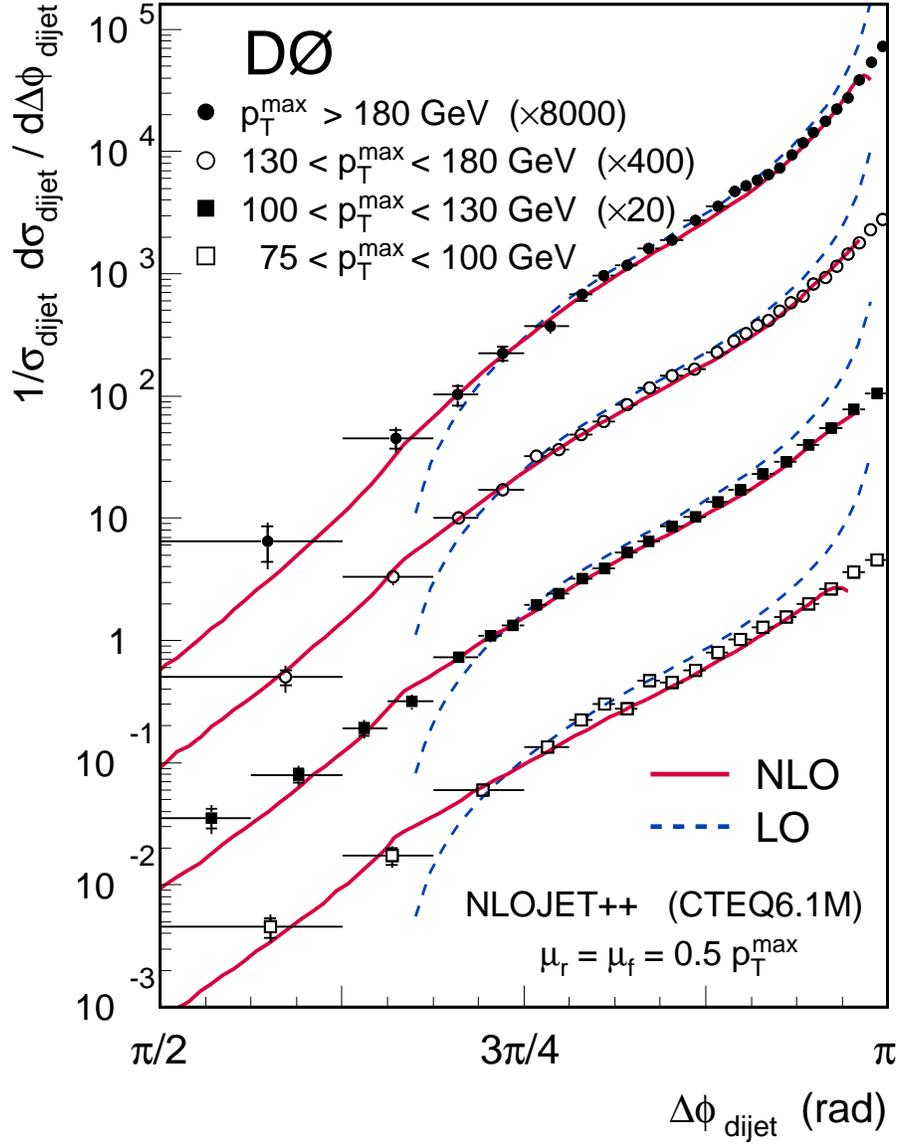,width=12cm}
\caption{Measurement of the difference in azimuthal angle between the two
leading jets in multijet events (D\O\ collaboration).}
\end{center}
\label{fig6}
\end{figure}

\begin{figure}
\begin{center}
\epsfig{file=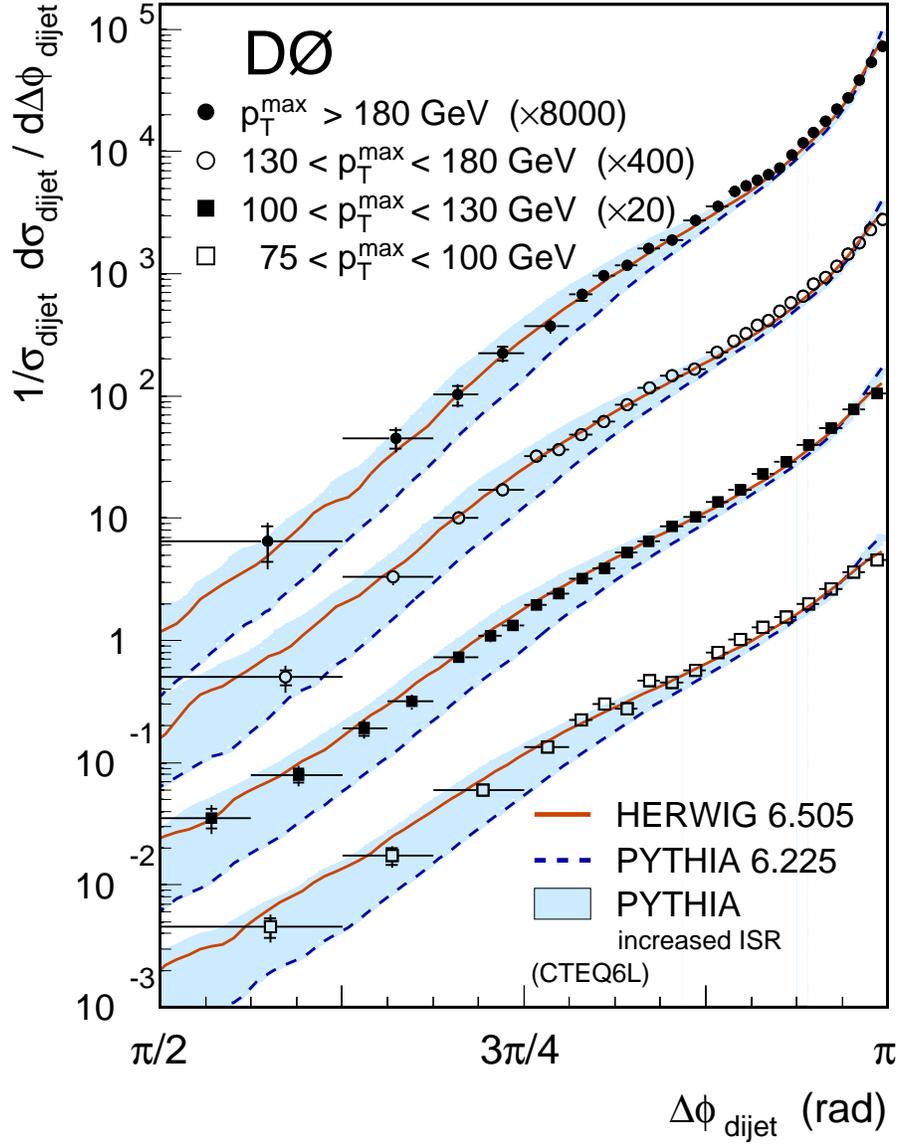,width=12cm}
\caption{Sensitivity of the measurement of the difference in azimuthal angle between the two
leading jets in multijet events to MC tuning (D\O\ collaboration)}
\end{center}
\label{fig7}
\end{figure} 

\subsection{Jet shape measurement}
\begin{figure}
\begin{center}
\begin{tabular}{cc}
\hspace{-1.7cm}
\epsfig{figure=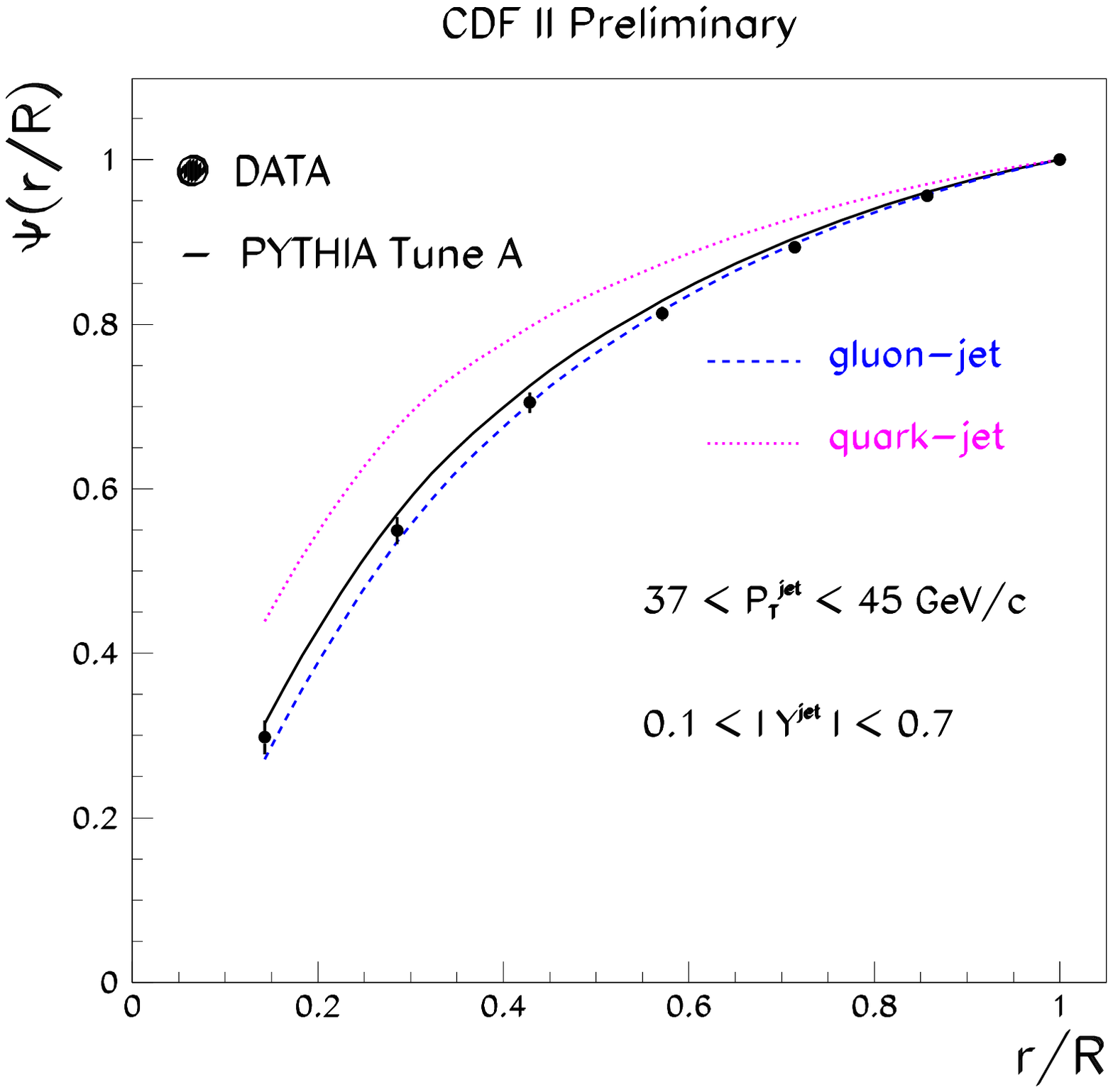,height=3.2in} &
\hspace{-1.cm}
\epsfig{figure=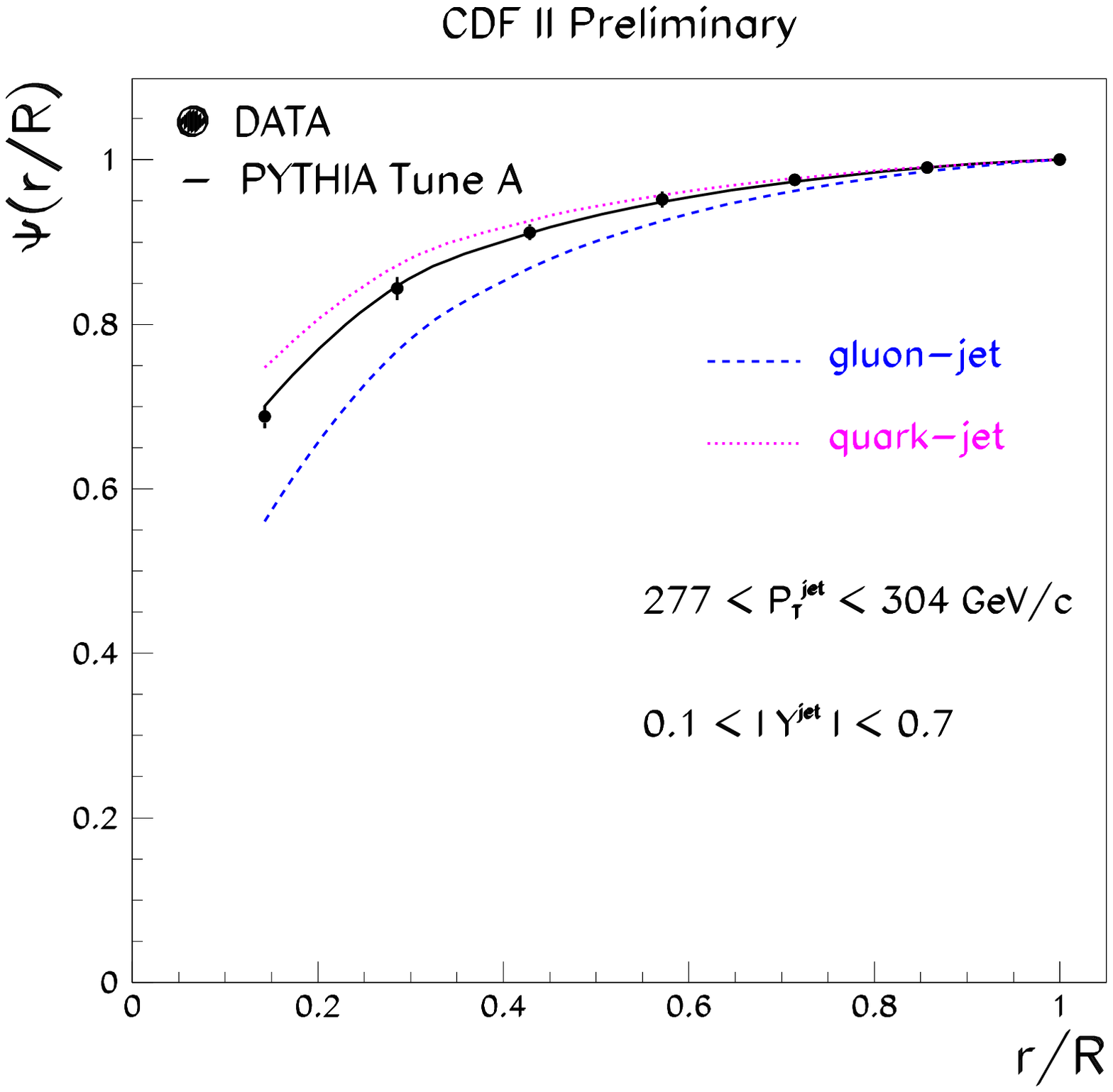,height=3.2in} \\
\end{tabular}
\end{center}
\caption{Jet shape distributions for two bins in jet transverse momentum.}
\label{fig8}
\end{figure}

The CDF collaboration performed another measurement sensitive to the gluon and
quark contents in the proton, as well as $\alpha_S$ and multi-gluon emission,
namely the jet shape measurement. The measurement consists in measuring
$\Psi(r)$ defined as follows:
\begin{eqnarray}
\Psi (r) = \frac{1}{N_{jets}} \Sigma_{jets} \frac{P_T(0,r)}{P_T^{jet}(0,R)},
\end{eqnarray}
where the summation runs over the number of jets in the event ($N_{jets}$),
and the jet radius is $R$.
$\Psi(r)$ is a measurement of the repartion of transverse energy within the jet.
Fig \ref{fig8} shows the jet shape distributions for two different bins in
jet transverse momentum, namely ($37 <p_T<45$ GeV) and ($277<p_T<304$ GeV) for central
jets ($0.1 < |y|<0.7$). 
The CDF measurement extends to more $p_T$ bins
\cite{cdfshape}. We also display in the same figure the expectations from the
PYTHIA \cite{pythia} Monte Carlo for gluon and quark jets. This measurement
allows to determine the proportion of quark and gluon jets as a function of
their transverse momentum. As expected, the lowest $p_T$ jets are gluon
process dominated whereas the higher $p_T$ jets are quark process dominated.

\subsection{Underlying events at the Tevatron}
Fig. 8 shows a typical jet event at the Tevatron. The upper
plots describes the hard scattering process where one observes the jet produced in
the event as well as the beam remnants. The lower plot displays what really
happens at the Tevatron (or later on at the LHC). In addition to the hard
scattering, we have initial and final state radiation which can produce
additional jets in the event, and additional partonic interactions not related
to the hard interaction (soft colour interactions can occur between the
spectator partons in addition to the hard interaction). This results in
additional energy measured in the detectors which are not related to the
partonic interaction. It is important to understand this phenomenon if one wants
to go back to parton level processes to measure the top mass, for instance.
To study these ``underlying events" (by opposition to the main hard scattering)
the CDF collaboration measured the energy emitted outside the dijet hemisphere
in clean back-to-back dijet events. For those events, one picks first the
direction of the leading jet in the events, and measures the energy in the
transverse region away from the leading jet. To avoid the particles included
in both jets, only the energy between 60 and 120 degrees in azimuthal angles
away from the leading jet is measured.
This energy is dominated by underlying events, or in other words,
by soft partonic interactions. The results were compared to the PYTHIA Monte
Carlo
\cite{pythia} and found to be in good agreement \cite{underlying} since PYTHIA
was already tuned to previous run I CDF data. It is important to note
that this tuning will have to be redone at the LHC aince it is not 
expected that
the energy of underlying events will be independent of the center-of-mass energy.

\begin{figure} 
\begin{center}
\begin{tabular}{c}
\epsfig{figure=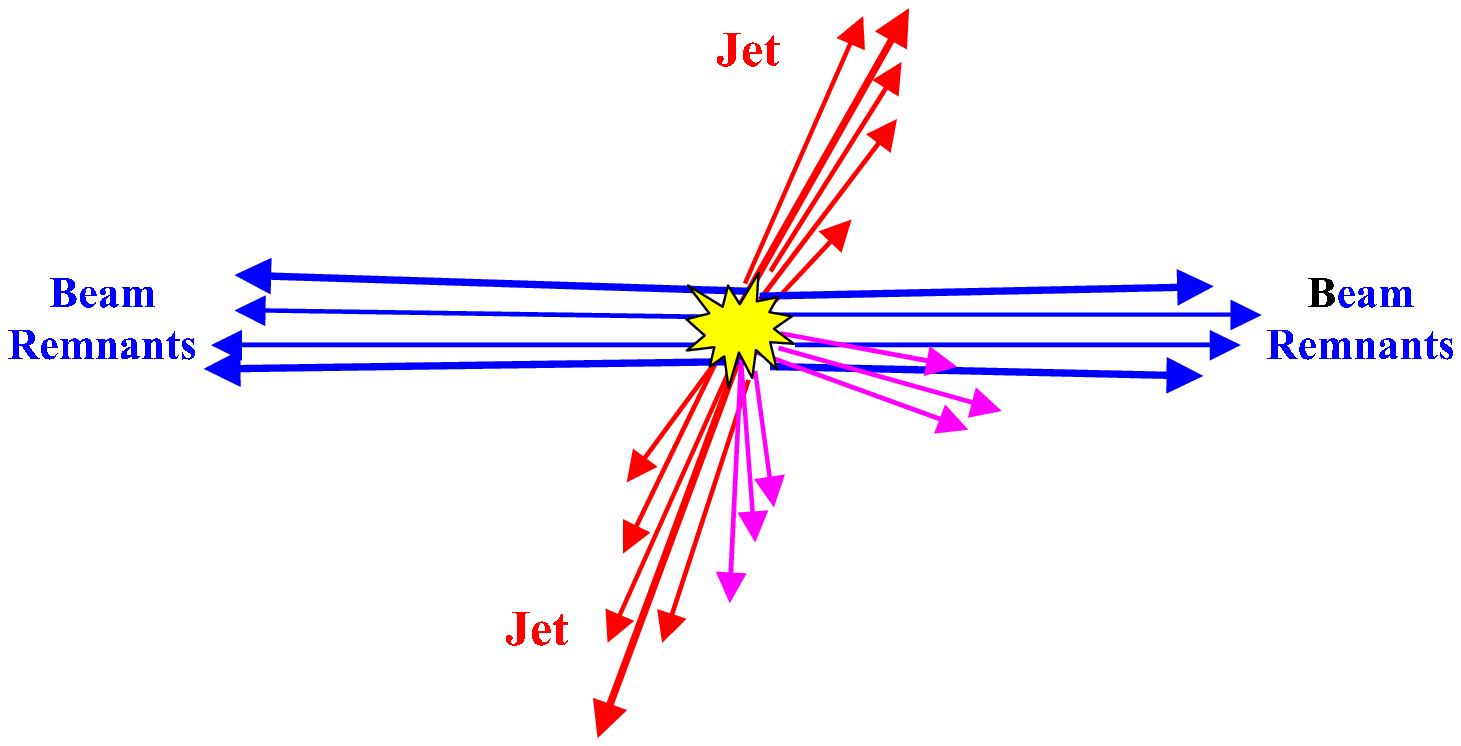,height=2.4in} \\
\epsfig{figure=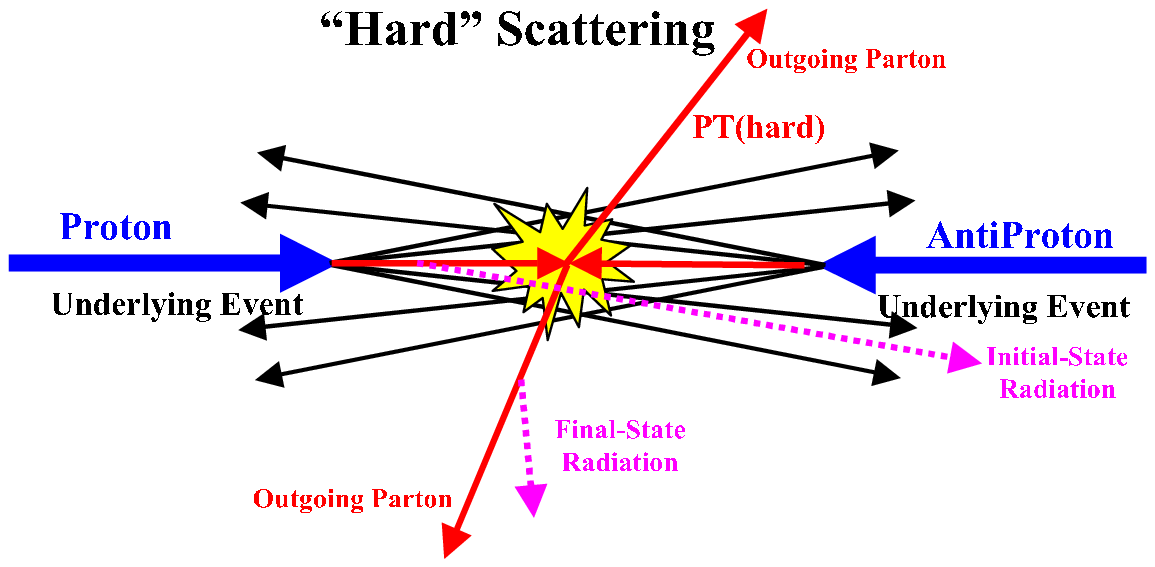,height=2.4in} 
\end{tabular}
\end{center}
\caption{Underlying events at the Tevatron.}
\label{fig9}
\end{figure} 

\section{Results on diffraction}

Diffractive events are of special interest since they show undestroyed protons
in the final state, and their mechanism is not yet fully understood. Mainly two
kinds of models exist to describe diffraction: the first model assumes the
existence of a colourless object, the Pomeron, which itself can be constituted
of quarks and gluons, and the other one assumes that
diffractive events are due to non perturbative string rearrangements 
in the final state (this happens at a much longer time scale than the hard
interaction, at the time scale of hadronisation). We distinguish between
single diffractive events and double pomeron exchanges which correspond to
diffractive events on the proton or antiproton side only or on both sides. 

\subsection{Structure of the pomeron}

Experimentally, there are two different ways to study diffractive events. The
first way is to detect directly events where there is no colour exchange between
the jet produced in the event and the proton in the final state, and to look for
a gap in rapidity in the forward region away from the proton direction.
The other way is to detect directly the proton in the final state in dedicated
detectors far away from the main detector in the tunnel called roman pot
detectors. The D\O\ and CDF collaborations installed this kind of detectors in
the tunnels. To describe diffractive events, one introduces two additional
kinematical variables: $\xi$ is the fraction 
of the proton momentum carried by the non coloured object (the Pomeron), and 
$\beta$ is the fraction of the pomeron momentum carried by the interacting
parton (quark or gluon) inside the Pomeron if we assume a partonic structure of
the Pomeron. By definition, $x_{bj} = \beta \times \xi$.
The CDF and D\O\ ``dipole" (close to the dipole magnets) roman pot
detectors are located at about 58 m away from the main detector in the 
outgoing antiproton direction and are sensitive to $t$ down to 0, and 
$0.02 < \xi <0.05$. The D\O\ collaboration installed in addition ``quadrupole"
roman pot detectors (close to the quadrupole magnets) in both outgoing proton
and antiproton directions located at about 23 and 33 meters away from the main
detector. These last detectors are sensitive to $|t|>0.5$ GeV$^2$, and
$10^{-3} < \xi < 3. 10^{-2}$. The commission of these detectors was recently
finished and new physics results are expected soon.

The percentage of single diffractive events was already measured by the D\O\ and CDF
collaborations in Run I and found to be about 1\% and depends on the exact
process considered. The amount of diffractive events at HERA, the $ep$ collider
located at DESY, Hamburg, is close to 10\%, which shows already that we cannot
obtain the Tevatron results directly from the HERA data, or in other words, that
there is no factorisation between $ep$ and $p \bar{p}$ colliders. This can be
due to additional soft interactions (soft gluon exchange) between partons in the
final state which kill the rapidity gap or destroy the proton in the final
state.

One important measurement on diffraction was performed in
Run I by the CDF collaboration \cite{cdfdiffrun1}. Using single diffractive events,
(an anti-proton was tagged in the roman pot detector), the CDF collaboration
was able to measure the gluon density in the Pomeron using dijet events.
The CDF data points and their error bands in yellow are shown in Fig. 9.
The results are compared directly to the expectations from the H1 diffractive
DGLAP QCD fits in red full line. We notice that there is a discrepancy in
normalisation by about a factor 10 between the CDF measurement and the HERA
expectations (this corresponds to the different in the
percentage of diffractive events between HERA and the Tevatron already
mentionned). However, in a large domain in $\beta$, the shape of the gluon
density is found to be similar which means that the same shape for the gluon
density can be used to describe HERA and Tevatron data, as well as probably LHC
data in the future. It is quite important to have precise measurements of the
gluon and quark densities inside the pomeron if one wants to make precise
predictions at the LHC \cite{diffinclus}.

Other measurements have been performed by the CDF collaboration \cite{cdfdiff}
concerning the tests of factorisation at the Tevatron. It was found that
factorisation holds almost in the full phase space at the Tevatron alone, and
that the same $x$ and $Q^2$ dependence has been found for inclusive or
diffractive jet production.

\begin{figure}
\begin{center}
\epsfig{file=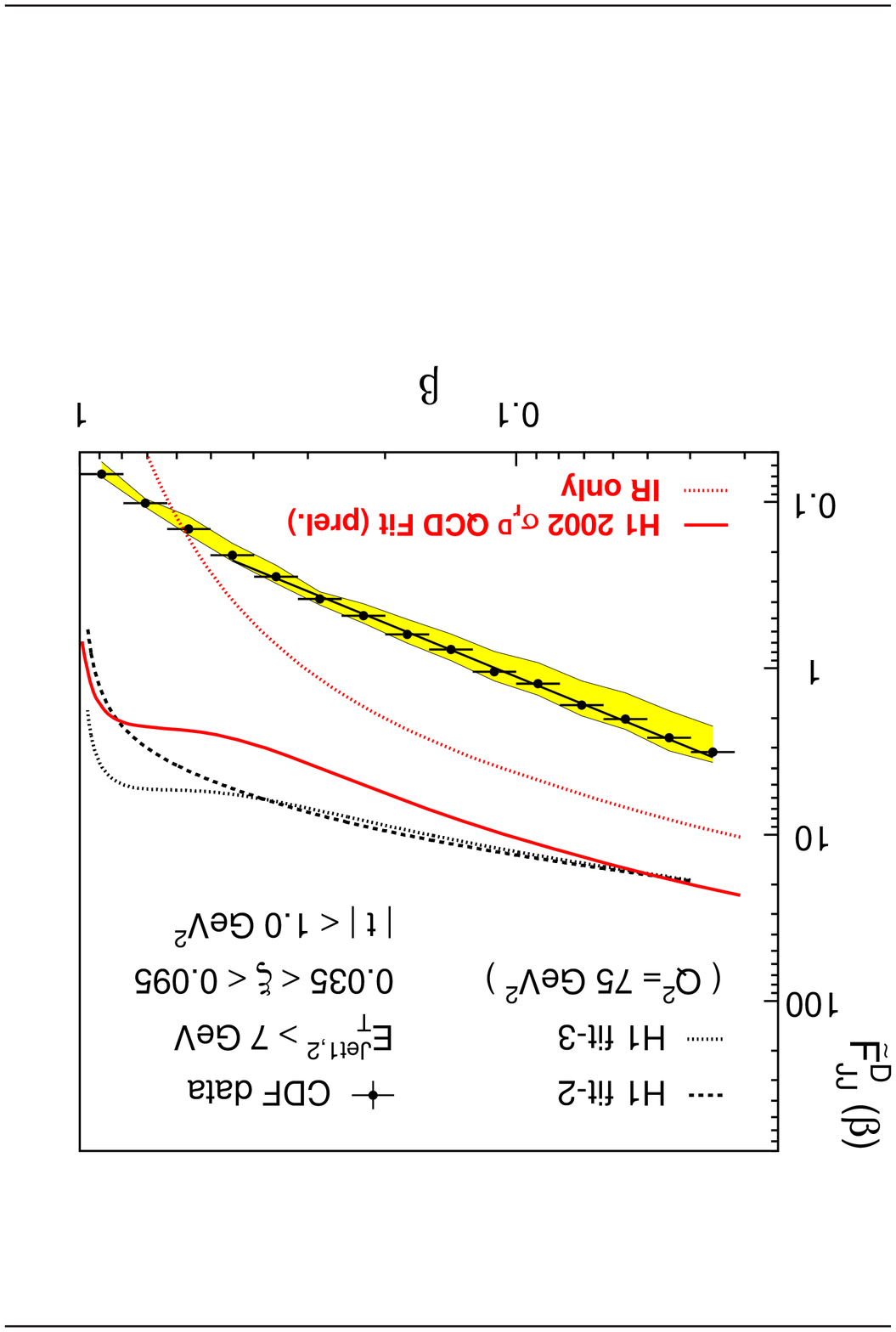,height=5.2in,angle=180,clip=true}
\caption{Comparison between the gluon density measured at the Tevatron 
(CDF data points) and the one measured at HERA (result of the H1 QCD fit in
full red line).}
\end{center}
\label{fig10}
\end{figure} 

\newpage

\subsection{Search for diffractive exclusive production}
Looking for the existence of exclusive events at the Tevatron is quite important
for the LHC. If exclusive events exist, it could be a way to look for
diffractive exclusive Higgs, top, or stop production at the LHC depending on the
production cross section \cite{diffexclus}, since it is possible to reconstruct
precisely the mass of the object produced diffractively using roman pot
detectors, using the so-called missing mass method, the total diffractive mass
produced being equal to $M= \sqrt{\xi_p \xi_{\bar{p}} S}$. The CDF collaboration started to
look for the eventual existence of exclusive events in the dijet channel.
The results are shown in Fig. 10 for a low luminosity of 26 pb$^{-1}$
(the actual accumulated luminosity by D\O\ and CDF is about 1 fb$^{-1}$ and we
can expect an update of these results very soon). The CDF data are divided in
three different samples corresponding to single diffraction (triangles), 
and double Pomeron exchange (empty and full circle points requiring a different
domain in rapidity for the gap: $5.5 < \eta < 7.5$ or $3.6 < \eta < 7.5$ for
empty and  full points respectively). The dijet mass fraction (the ratio of the
dijet mass by the total diffractive mass in the event) is displayed in Fig.
10. Exclusive events are expected to appear at large dijet mass
fraction since the full energy is used to produce dijets (there is no loss of
energy in pomeron remnants). No enhancement is observed at high dijet mass
fraction which is compatible with the tail of the inclusive distribution, but
the cross section for exclusive production is expected to be small.
It will be quite interesting to see the results with higher luminosity.
Other methods can also be developped to look for exclusive events like measuring
the correlation between $\log 1/\xi$ and the size of the rapidity gap which is
larger for exclusive events, the ratio of the dilepton to diphoton cross sections
which should show an enhancement at high diphoton-dilepton mass if exclusive
events exist, or the ratio between b and light jet diffractive production
\cite{diffexclusb}. 

Another method is to look for diffractive $\chi_C$ production. Unfortunately,
the acceptance for such low mass objects to be detected in roman pot detectors
is small and the selection requires the existence of rapidity gaps. The
diffractive mass has to be computed using the central calorimeter without
benefitting from the good resolution of the roman pot detectors. The CDF
collaboration looked for $\chi_C^0$ decaying into dimuon and a photon, and no
further activity in the central detector was requested 
to ensure the exclusiveness of the process. A few
exclusive candidate events were found but it is difficult to determine precisely
the cosmic contamination \cite{chiccdf}. 

\begin{figure}
\begin{center}
\epsfig{file=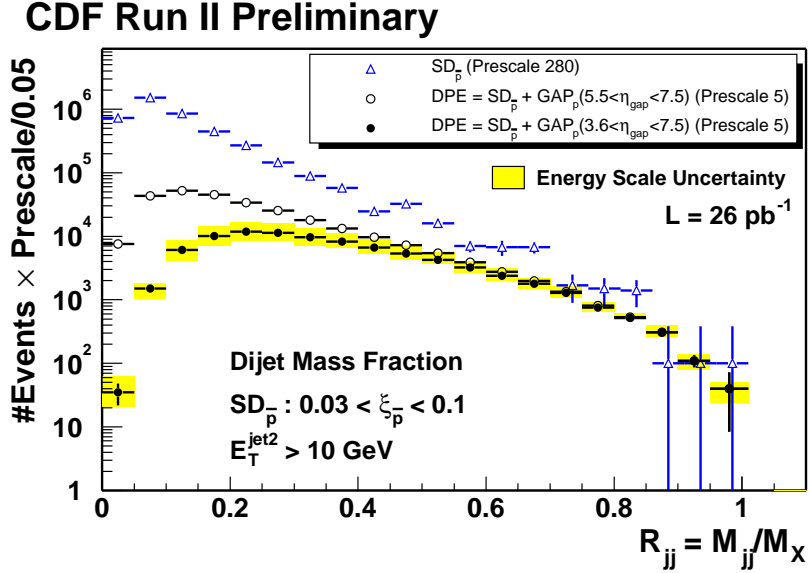,width=12cm}
\caption{Dijet mass fraction measured in the CDF detector for double Pomeron
exchange.}
\end{center}
\label{fig10}
\end{figure}

\section{Top physics}
Top physics is one of the hottest subjects at the Tevatron, which is the only
place where the top quark can be studied before the start of LHC. The top
quark was indeed discovered at the Tevatron Run I in 1995 by the D\O\ and
CDF collaborations. Compared to the other quarks, it has a much
higher mass (its mass is about 174 GeV which is 40 times the bottom quark mass).
Due to its mass, the top quark life time is very short (about 10$^{-25}$s), and
the top quark decays before hadronisation. In Fig.
12, the schematic production and decay of a typical $t
\bar{t}$ event is displayed. The production cross section at the Tevatron is of the order
of 6 pb, 85\% of which are produced via a $q \bar{q}$ interaction, and 15\% via
a $gg$ one. The top quark decays into a $W$ and a $b$ quark in 100\%
of the cases since $V_{tb}$ is much greater than $V_{ts}, V_{td}$. The $W$ can
decay either leptonically as indicated in the figure or into 2 jets (quarks
$u \bar{d}$). A typical topology to look for $t \bar{t}$ events is a multijet
event (6 jets, 2 can be b-tagged), or a multi jet and lepton event with missing
transverse energy coming from the $W$ decay.

\begin{figure}
\begin{center}
\epsfig{file=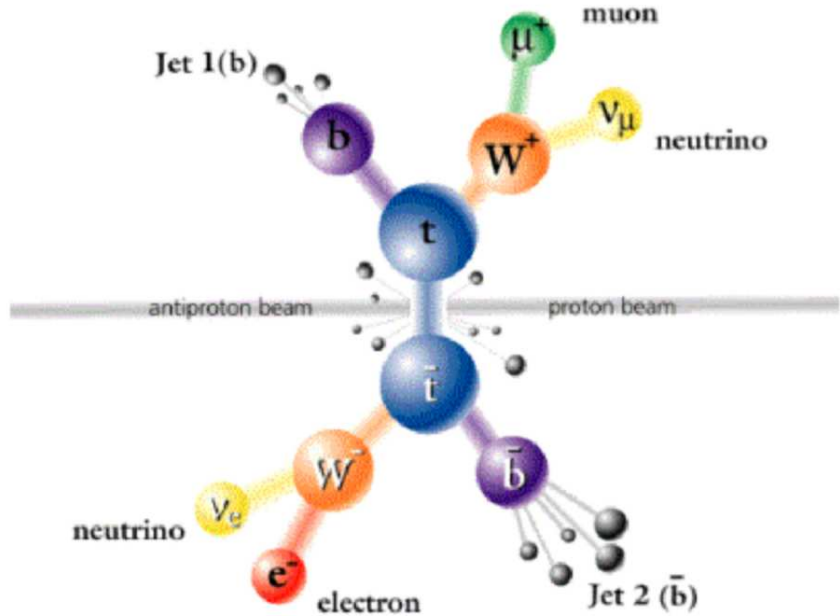,width=12cm}
\caption{Scheme of a $t \bar{t}$ event.}
\end{center}
\label{fig11}
\end{figure}

\subsection{Measurement of the top quark mass}
The measurement of the top quark mass is a fundamental test of the Standard
Model. The radiative corrections to the Standard Model predictions of
electroweak measurements are dominated by the value of the top mass, and a
precise measurement of the top mass is needed to constrain the electroweak tests of the
Standard Model and the Higgs boson mass. The measurement of the top mass depends
first on the identification of the $t \bar{t}$ events by requiring a leptonic,
multijet (at least 4) event, and missing transverse energy. The background to this
topology can be further reduced requiring some jets to be b-tagged. The mass
measurement is also very sensitive the determination of the jet energy scale.
One of the easiest methods to determine the top mass is to use the template
method. The basic idea is to compute a $\chi^2$ between data and
Monte Carlo simulations assuming different values of the top mass. In fact, the
method is slightly more complicate: it is for instance possible to constrain the
jet energy scale by constraining the measurement of the $W$ mass in data to be
in agreement with the world average since the $W$ mass is already known
precisely. The different Run II measurements of the top quark mass (at the time
of the summer school) \cite{topmass} are given in
Fig. 13 for the D\O\ and CDF collaborations. By comparison, the Run I
average was 178 $\pm$ 4.3 GeV  and the best single top mass measurement was
performed in the lepton and jet channel by the D\O\ collaboration \cite{topmassd0}
(180.1 $\pm$ 5.3 GeV). A precision on the top mass a bit higher than 1 GeV is expected
by the end of Run II at the Tevatron. The new Run I top mass led to
the prediction of the Higgs boson
mass of (114 $+$ 69 $-$ 45) GeV using electroweak fits. Reducing the uncertainty
on the top mass will allow to reduce its large uncertainty.

\begin{figure}
\begin{center}
\epsfig{file=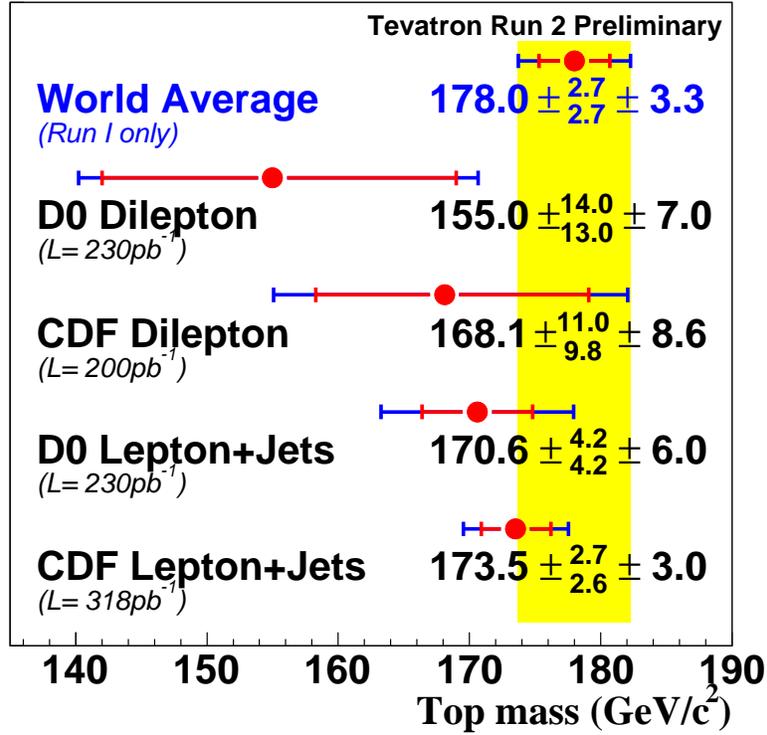,width=12cm}
\caption{Measurement of the top quark mass.}
\end{center}
\label{fig11}
\end{figure}

\subsection{Measurement of the $t \bar{t}$ production cross section}
The analysis of the $t \bar{t}$ events described in the previous paragraph leads
directly to a measurement of the $t \bar{t}$ production cross section and can be
compared directly to the prediction of the Standard Model. Many different
methods (dilepton, lepton and jet, multi jet channels) are used by the CDF and
D\O\ collaborations \cite{topcross}. The combined result for the CDF 
collaboration is given in Fig. 14.

\begin{figure}
\begin{center}
\epsfig{file=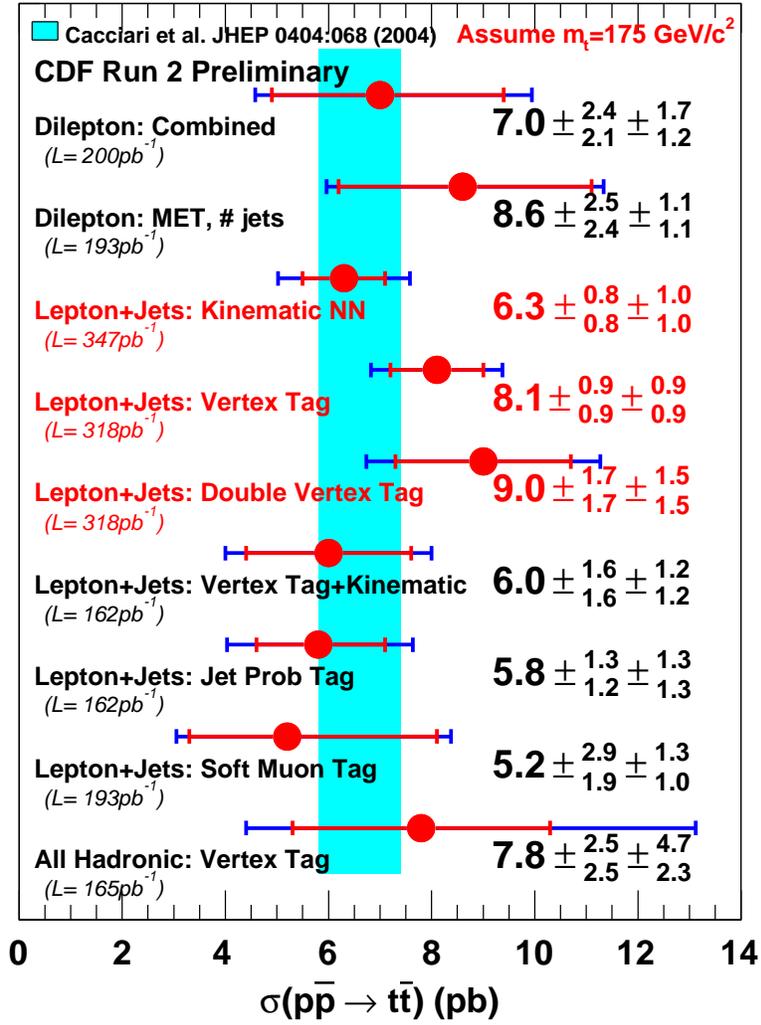,width=12cm}
\caption{$t \bar{t}$ production cross section (CDF collaboration).
The D\O\ collaboration shows similar results.}
\end{center}
\label{fig12}
\end{figure}

\subsection{Search for single top production}
Another way to produce the top quark predicted by the Standard Model 
is the electroweak single production, where the top quark is produced via a $W$.
This process has not yet been observed, but a limit at 95\% CL was set by the D\O\
collaboration on the production cross section at 6.4 pb in the $s$-channel and
5 pb in the $t$-channel \cite{singletop}. The limit is now close to the cross section predicted
by the Standard Model and an observation could come soon. The advantage of that
process is to study the CKM matrix element $V_{tb}$, the top width and the
$Wtb$ coupling.

\section{Electroweak physics}

\subsection{Measurement of $W$ and $Z$ production cross sections}
$Z$ and $W$ bosons can be produced directly by quark interactions at the
Tevatron. To obtain a lower background, one measures the $W$ and $Z$ cross 
sections when the $Z$ or the $W$ decays into dileptons or lepton and neutrino
respectively. The CDF and D\O\ results are given in Fig. 15 and Fig.
16 for $Z$ and $W$ production respectively
\cite{wz}. The results obtained in Run
I (center-of-mass energy of 1.8 TeV) are displayed together with the new Run II
results (center-of-mass energy of 1.96 TeV) and compared with the Standard 
Model expectation (full line). The data points are not put all at either 1.8 
or 1.96 TeV to be able to distinguish between them. The different leptonic
decays of the $Z$ or $W$ are shown (electron, muon or tau) and we also note the
good agreement between the measurements. 

Another important measurement to be performed at the Tevatron is the $W$ mass.
Some update on this subject are expected in the near future. The measurement
requires a very good understanding of the systematics to be able to obtain a
world competitive measurement.

\begin{figure}
\begin{center}
\epsfig{file=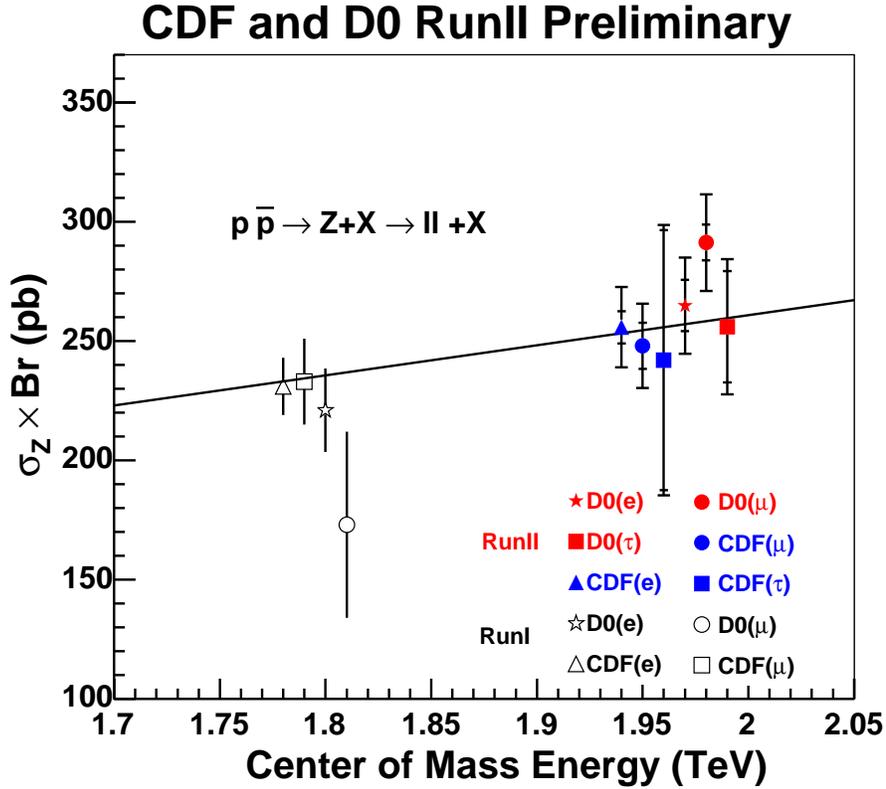,width=12cm}
\caption{$Z$ production cross section.}
\end{center}
\label{fig13}
\end{figure}

\begin{figure}
\begin{center}
\epsfig{file=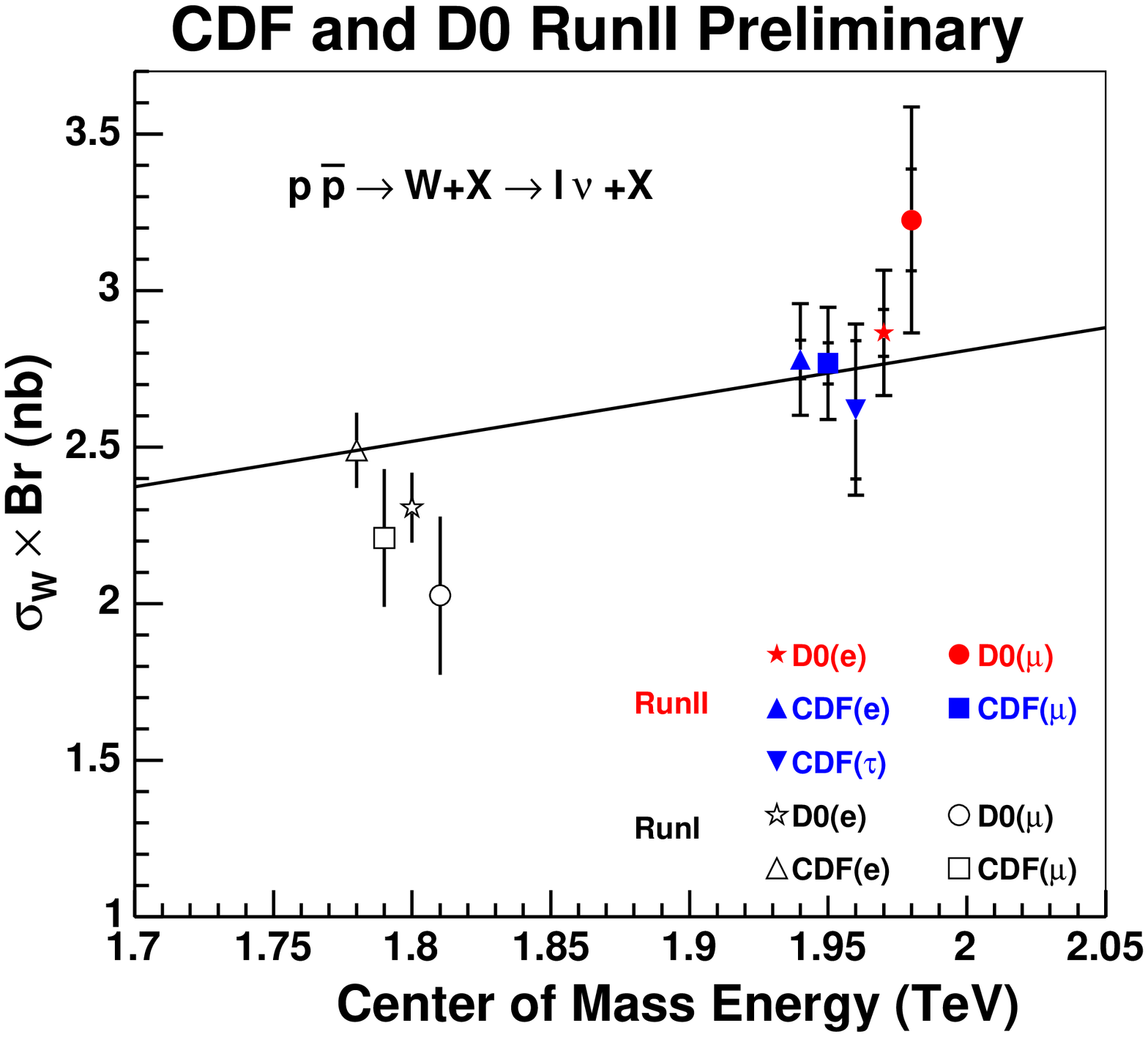,width=12cm}
\caption{$W$ production cross section.}
\end{center}
\label{fig14}
\end{figure} 

\subsection{$W$ asymmetries}
The $W$ asymmetries have been measured by the CDF collaboration. The advantage
of this measurement is that it is sensitive to $u$ and $d$ contents of the
proton. In average, $u$ quarks carry more proton momentum than $d$ quarks.
As a consequence, the rapidity distribution for $W^+$ is different from the
one for $W^-$. Namely, $W^+$ which are produced mainly by $u$ and $\bar{d}$ interaction
receive a boost in the $u$ direction, and $W^-$ which are produced by $d$ and
$\bar{u}$ in the $\bar{u}$ direction. This explains why the rapidity
distribution for $W^+$ (respectively $W^-$) has the tendency to be shifted
towards positive (respectively negative) values of rapidity. The CDF 
collaboration measured
the $W$ asymmetries defined as follows:
\begin{eqnarray}
A(y)= \frac{d \sigma (W^+)/dy - d \sigma (W^-)/dy}
{d \sigma (W^+)/dy + d \sigma (W^-)/dy} \sim \frac{d}{u} 
\end{eqnarray}
which gives a direct access to the ratio of $d$ and $u$ quark densities.
The result is shown in Fig. 17
for a transverse energy bin between 35 and 45 GeV as a function of $W$ rapidity. 
The expectations from the
CTEQ and MRS distributions are also given \cite{asymmetry}. We see that the main
differences occur at high rapidity. With more 
accumulated luminosities, it will possible to perform the same measurement at
higher energy which will give more sensitivity on the quark densities.

\begin{figure}
\begin{center}
\epsfig{file=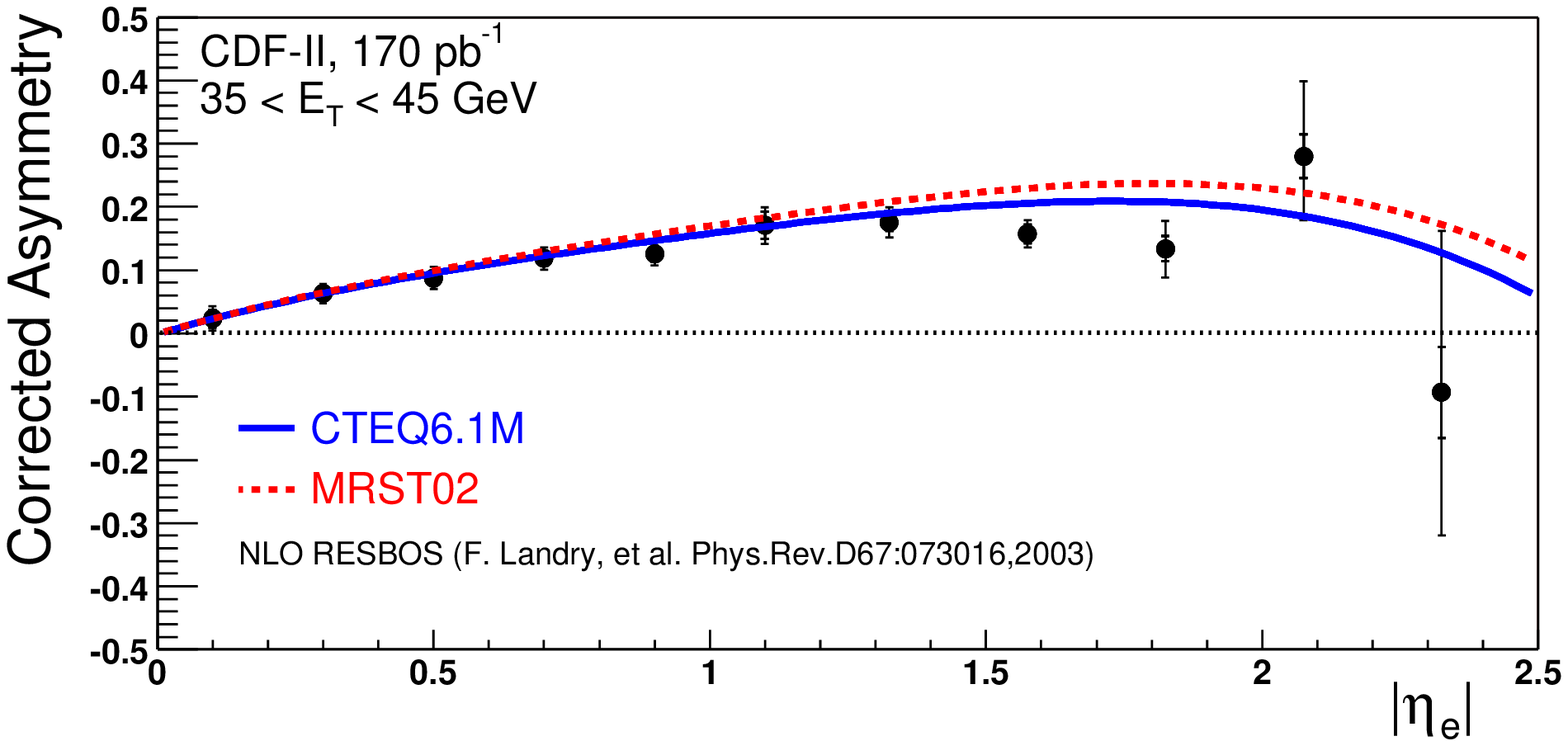,width=12cm}
\caption{$W$ asymmetries (CDF collaboration).}
\end{center}
\label{fig15}
\end{figure}

\section{B physics}

Many results have been published already by the CDF and D\O\ collaborations
concerning B physics. Due to the lack of time, we will cover only a few topics.
Other results can be found on the web pages of the collaborations
\cite{bphysics}.

A general plot showing the resonances appearing in the dimuon systems can
already give a feeling on the excellent mass resolution obtained by the D\O\ and
CDF detectors due to their tracking and silicon detectors. Fig. 18
displays the $\omega$, $\Phi$, $J/\Psi$, $\Psi'$ and $\Upsilon$ resonances
observed by the D\O\ collaboration in the dimuon system. Other resonances such
as $B^+$, $\Phi$, or $\Lambda_b$, ... have also been studied by the D\O\ and CDF
collaborations \cite{bphysics}.

The D\O\ collaboration also observed the $X(3872)$ resonance \cite{x3872} in the 
$J / \Psi ~ \pi^+ \pi^-$ channel as it is shown in Fig. 19. The mass
difference between $X(3872)$ and $J / \Psi$ has been found to be 
774.9 $\pm 3.1 (stat.) \pm 3.0  (syst.) $MeV.

\begin{figure}
\begin{center}
\epsfig{file=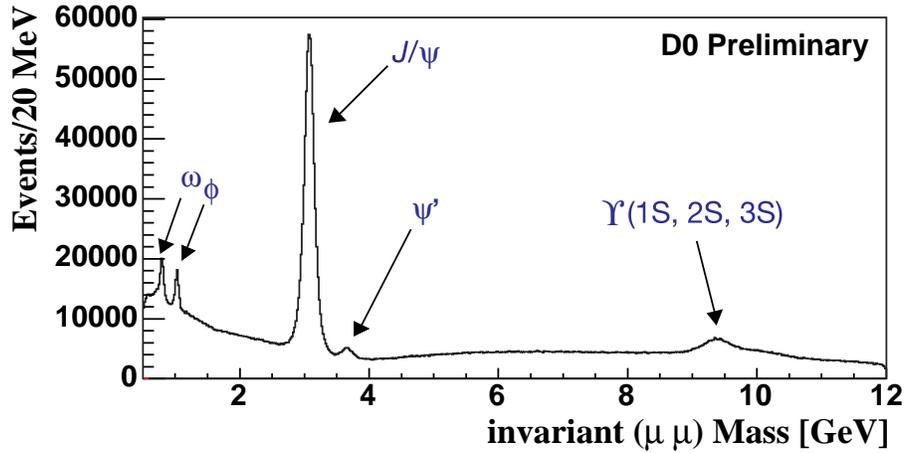,width=12cm}
\caption{Dimuon resonances observed by the D\O\ collaboration.}
\end{center}
\label{fig15}
\end{figure}

\begin{figure}
\begin{center}
\epsfig{file=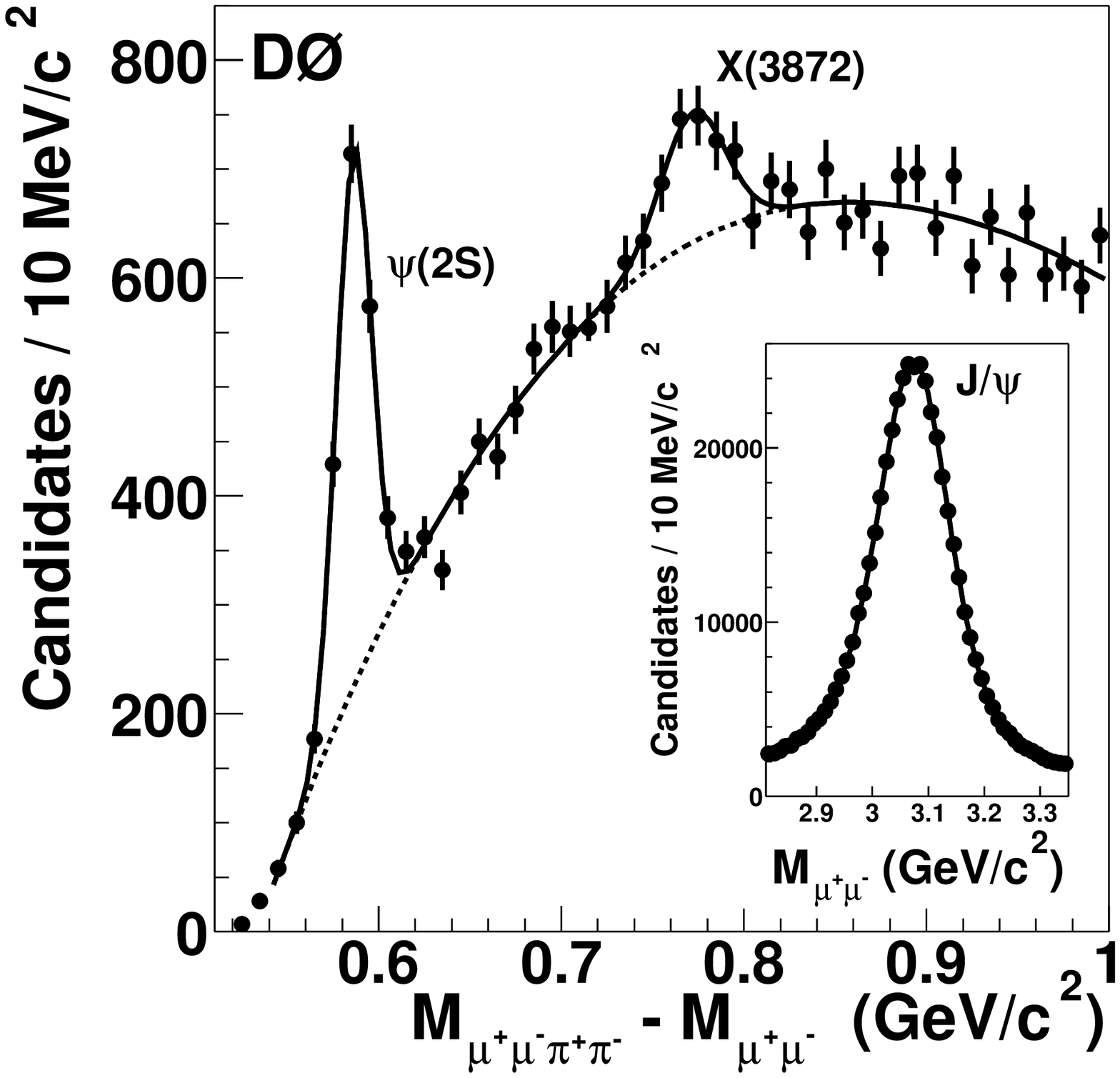,width=12cm}
\caption{Observation of $X(3872)$ by the D\O\ collaboration.}
\end{center}
\label{fig16}
\end{figure}

\section{New phenomena}
The new phenomena studies are done by the D\O\ and CDF collaborations mainly in
the SUSY framework. We defined the so-called $R$-parity which is  
$(-1)^{2j+B+L}$ where $j$, $B$ and $L $ stand for spin,
baryon and lepton numbers. Standard Model (respectively SUSY) particles 
show $R=1$ (respectively $R=\pm 1$). The experimental signatures to look for SUSY
particles are different if $R$-parity is conserved or violated. When
$R$ parity is conserved, SUSY particles are produced in pairs, and they
decay into the lightest SUSY particle (LSP) which
escapes undetected. Experimentally, this induces some missing transverse energy
which can be detected. On the contrary,when $R$ parity is not conserved,
the LSP decays, and the experimental signature is an event with multi-lepton,
multi-jets, with little missing transverse energy, and the process often
includes lepton flavour violating decays.

Before describing the search for new phenomena, let us give some feelings about
the cross section we are concerned with. Typical jet production cross section at
the Tevatron are of the order of 10$^{12}$ fb (10$^{11}$ fb for $b$-jets),
whereas the $W$ and top typical cross sections are in the order 10$^7$ and
a few 10$^3$ as we mentioned in previous paragraphs. The present limits on
SUSY particle production cross section lay in the region of 10$^4$ fb for squark
production and a few tenths of fb for sleptons. We already see that the main
problems of new phenomena analysis will be to get rid of the huge background
without losing too many new phenomena events since they are expected to be rare.

We will not give here a complete exhaustive list of all new phenomena results
but rather focus on three particular ones. All results from the D\O\ and CDF
collaborations can be found on their web pages \cite{susyweb}.  

\subsection{Squarks and gluinos}
Squarks and gluinos can be produced directly by pairs at the Tevatron via a $q \bar{q}$
interaction. The squarks decay into the LSP (assumed to be the
$\tilde{\chi}_1^0$) and a quark. The topology for squark pair production will be 2
jets and missing transverse energy. Similarly, the topology for squark
gluino or gluino pair production is respectively two jets and missing transverse
energy or three jets and missing transverse energy. No signal has been found in
this channel and the limit has been obtained by the D\O\ collaboration in the
squark-gluino mass plane \cite{squarksd0} for 310 pb$^{-1}$ as shown in Fig.
20. The previous limits from LEP and Tevatron Run I are also displayed
on the figure.

\begin{figure}
\begin{center}
\epsfig{file=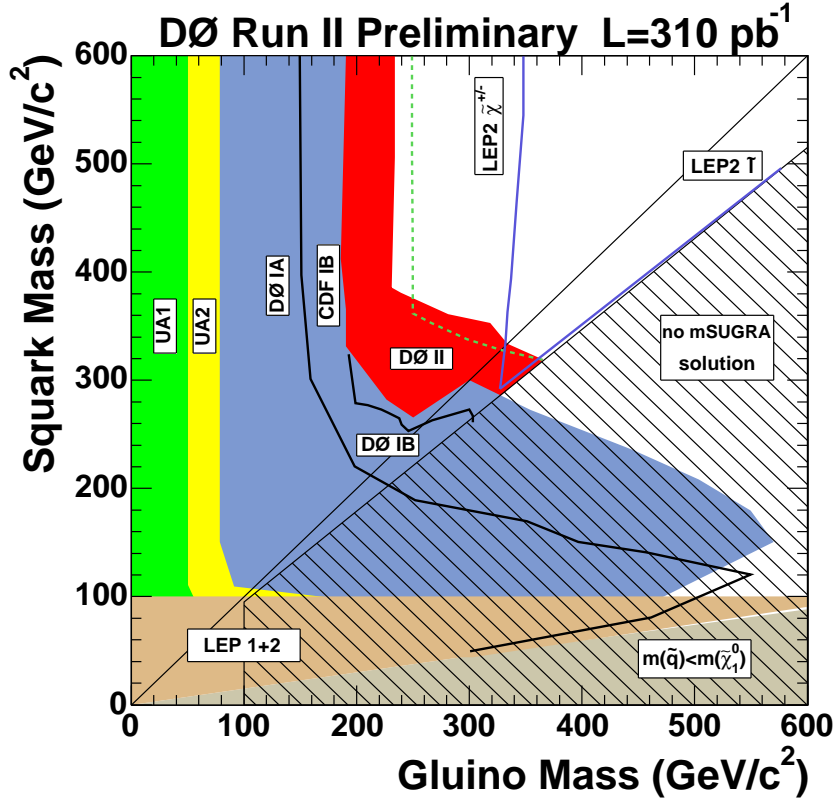,width=12cm}
\caption{Limits in the squark-gluino mass plane obtained by the D\O\
collaboration}
\end{center}
\label{fig17}
\end{figure}

\subsection{Stop production in mSUGRA}
The CDF collaboration studied the production of stop pair in minimal
supergravity (mSUGRA) scenario. Stops are produced in pair as in the previous
squark production. The stop (assumed to be the next lightest supersymmetric
particle) is assumed to decay into $c \tilde{\chi}_1^0$ where the $\tilde{\chi}_1^0$ is
assumed to be the LSP. The selection is thus to require two reconstructed 
jets coming from the $c$ quark and missing transverse energy from the LSP.
The study is made for different mass values of the LSP, and as an example, we
show the results for a LSP mass of 40 GeV in Fig. 21. The CDF limit is
displayed in full and the stop production cross section in dashed line for the
CTEQ5M parametrisation \cite{cdfstop}.

\begin{figure}
\begin{center}
\epsfig{file=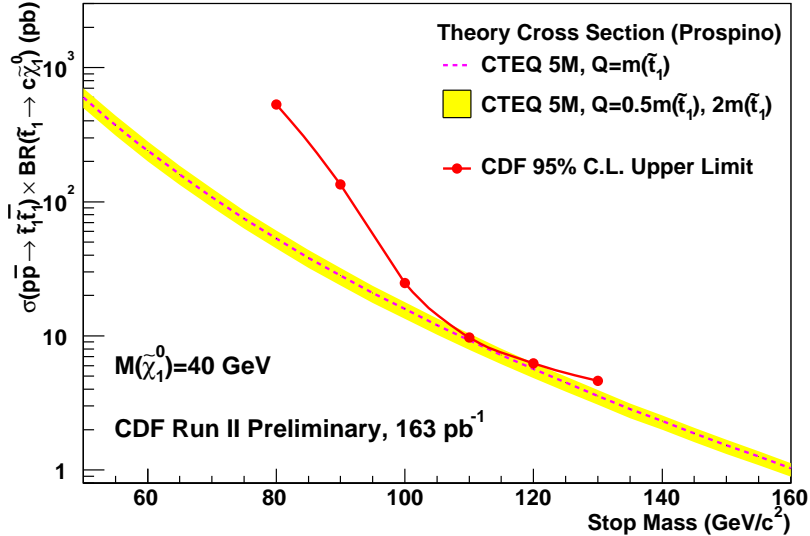,width=12cm}
\caption{Limits on the stop pair production cross section times branching ratio
by the CDF collaboration.}
\end{center}
\label{fig18}
\end{figure} 

\subsection{Resonant sparticle production with violated $R$-parity}
When $R$-parity is violated, it is possible to produce sparticles in the
$s$-channel in a resonant mode \cite{resonant}. For instance, it is possible to
produce smuons from $d$ and $\bar{u}$ quarks and the so-called $\lambda'_{211}$
coupling. In the same way, the LSP can decay via another $R$-parity violating
coupling. New limits have been established by the D\O\ collaboration for
resonant sparticle production for the $\lambda'_{211}$ coupling for different
neutralino and slepton masses. As an example, we display in Fig. 22
the limits on the  $\lambda'_{211}$ coupling as a function of the neutralino 
mass for a fixed slepton mass of 200 GeV \cite{rpvd0}, the Run I result being
indicated for reference.

\begin{figure}
\begin{center}
\epsfig{file=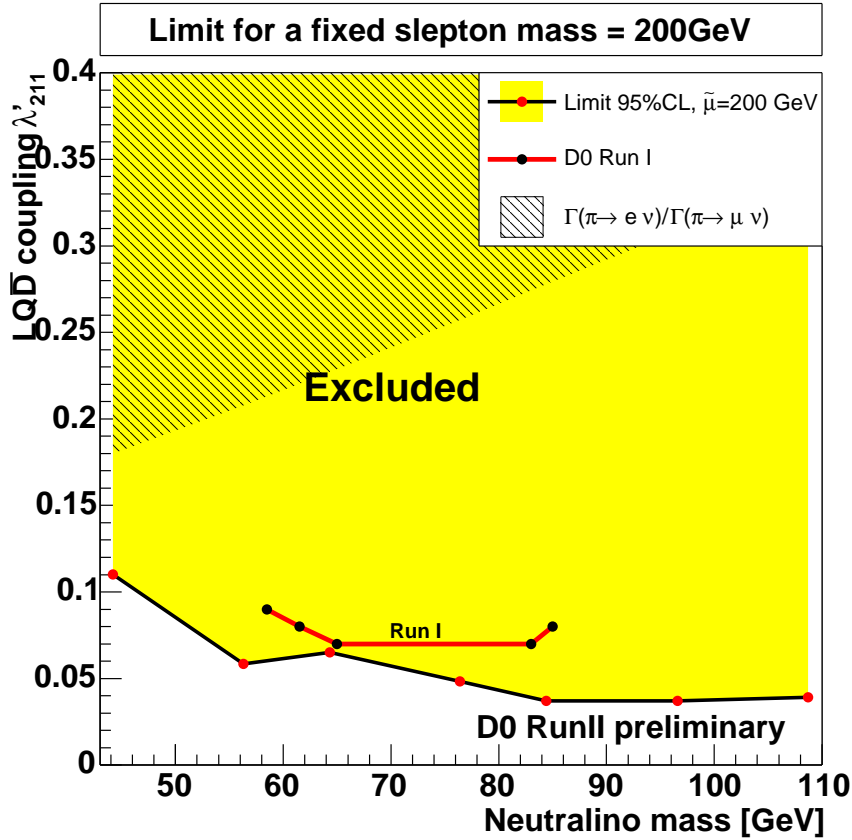,width=12cm}
\caption{Limits on the $\lambda'_{211}$ coupling as a function of the neutralino 
mass for a fixed slepton mass of 200 GeV from the D\O\ collaboration.}
\end{center}
\label{fig19}
\end{figure}

\subsection{Search for Higgs boson}
A hot but difficult topic for the Tevatron is the search for neutral Higgs
bosons. Predictions have been made on the sensitivity to look for Higgs bosons
in the next years when luminosity increases and are given in Fig. 23.
These results will strongly depend on the detector performances since the
background is very high in all channels and the search for Higgs boson quite
challenging. The large error band shows the expectations for a 5$\sigma$
dicovery, 3$\sigma$ evidence, and a 95\% CL limit as a function of the Higgs
mass from an analysis of the Higgs sensitivity study working group \cite{higgs} (the smaller band 
shows the previous results). However these results do not include systematic
errors but only statistical ones and are thus optimistic.

\begin{figure}
\begin{center}
\epsfig{file=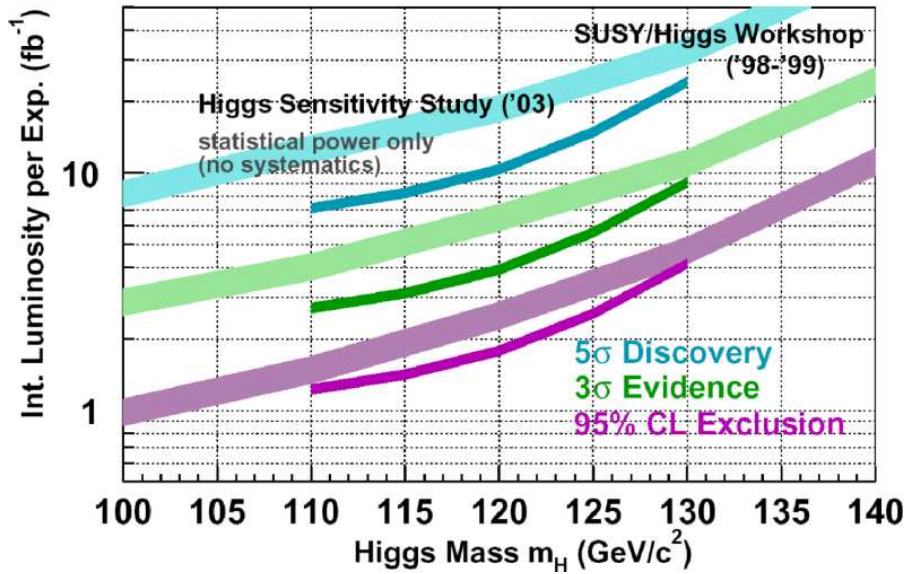,width=12cm}
\caption{Prospects to search for Higgs bosons at the Tevatron.}
\end{center}
\label{fig20}
\end{figure} 

\section{Conclusion}
In these lectures, we have discussed many preliminary results from the Tevatron
on QCD, diffraction, electroweak, top and B physics, and new phenomena. Much
progress is expected in the future with the increase of luminosity (this will
benefit directly to new phenomena studies and the search for Higgs bosons) and
a better understanding of systematics which are often dominated by the
uncertainty on jet energy scale (QCD cross section measurements and constraint
on the parton distributions, electroweak physics and the $W$ mass measurement,
top physics and the top mass measurement allowing to constrain further the
standard model and the mass of the Higgs boson).

\section*{Acknowledgments}
The author thanks the organisers of the Zakopane Summer School for
financial support and Jochen Cammin and Robi Peschanski for a careful reading of
the manuscript.


\end{document}